\documentclass[prd,twocolumn,superscriptaddress,altaffilletter,nofootinbib]{revtex4}

\usepackage{cancel}
\usepackage{graphicx}
\usepackage{amsmath}
\usepackage{amssymb}
\usepackage{graphicx,epsfig}
\newcommand{\be}{\begin{equation}}
\newcommand{\ee}{\end{equation}}
\newcommand{\bea}{\begin{eqnarray}}
\newcommand{\eea}{\end{eqnarray}}
\newcommand{\der}{\partial}
\newcommand{\vphi}{\varphi}

\begin{document}



\title{Revisiting local-scale invariant gravitational theory}

\author{Israel Quiros}\email{iquiros@fisica.ugto.mx}\affiliation{Dpto. Ingenier\'ia Civil, Divisi\'on de Ingenier\'ia, Universidad de Guanajuato, Campus Guanajuato, Gto., M\'exico.}

\date{\today}

\begin{abstract} We revisit the conformally coupled scalar gravitational theory. This is the simplest local-scale invariant theory of gravity which is linear in the curvature scalar. We demonstrate that, if incorporate local-scale symmetry into the variational procedure, it is not required that the trace of the stress-energy tensor of the matter fields vanished for this symmetry to be preserved. The relevance of this result for the understanding of local-scale symmetry along with its physical consequences, is discussed.\end{abstract}


\maketitle




\section{Introduction}
\label{sect-intro}


The most widely used hypothesis on local-scale symmetry (LSS) is that this must be a broken symmetry in nature \cite{deser-1970, callan-1970, anderson-1971, fujii-1974, freund-1974, englert-1975, englert-1976, bekenstein-1980}. This hypothesis can be formulated in other equivalent ways as, for instance, that LSS is not compatible with dimensionful parameters such as masses and other physical constants. According to the latter formulation, the breakdown of the LSS is a consequence of the appearance of physical scales such as the Planck mass $M_\text{pl}$. Take, as an example, the conformally coupled scalar gravitational (CCSG) theory, which is given by the following Lagrangian \cite{callan-1970, deser-1970}:

\begin{align} {\cal L}_\text{ccsg}=\frac{\sqrt{-g}}{2}\left[\frac{1}{6}\phi^2 R+(\der\phi)^2-\frac{\Lambda}{4!}\phi^4\right],\label{ccs-lag}\end{align} where $R$ is the curvature scalar of Riemann space $V_4$, $\phi$ is the gauge scalar field with mass dimension (also known as compensator field) and $\Lambda$ is a dimensionless constant.\footnote{Many aspects of the theory \eqref{ccs-lag}, have been investigated over the years \cite{callan-1970, deser-1970, englert-1976, narlikar-1977, duncan-1977, anderson-1985, hans-1988, page-1991, ng-1991, madsen-1993, barcelo-1999, kallosh-2013-a, kallosh-2013-b, kallosh-2013-c, bars-2014-a, bars-2014-b, jackiw-2015, bambi-2017}. In recent dates the interest in the CCSG theory has increased in connection with non-Riemannian geometric backgrounds and local-scale symmetry \cite{ghilen-2019, ghilen-ejpc-2020, ghilen-2020, pgs-2021, ghilen-2022, harko-epjc-2022, oda-2022, babichev-2023, ghilen-2023, harko-prd-2023, harko-prd-2024-1, harko-pdu-2024, harko-prd-2024, quiros-prd-2023, rodrigo-arxiv}. Solution-generating techniques have been used to generate nonsingular spacetimes \cite{narlikar-1977}, as well as families of stationary solutions to several systems, including the CCSG theory \cite{duncan-1977}. Several new exact solutions of CCSG theory have been found which include a class of traversable wormholes and vacuum solutions in arbitrary dimensions \cite{barcelo-1999, babichev-2023}. A static spherically symmetric solution of the vacuum EOM derived from \eqref{ccs-lag} was found in \cite{harko-epjc-2022} (see also \cite{harko-prd-2023, harko-prd-2024-1, harko-pdu-2024, harko-prd-2024}).} We use the following nomenclature: $(\der\phi)^2\equiv g^{\mu\nu}\der_\mu\phi\der_\nu\phi$. The wrong sign of the kinetic energy density term for $\phi$-field is harmless since, due to LSS the scalar field can be ``gauged away.'' The gravitational Lagrangian \eqref{ccs-lag}, as well as the derived equations of motion (EOM) are invariant under the local-scale transformations (LST), also known as either ``conformal transformations,'' ``Weyl rescalings'' or ``transformations of units'' \cite{anderson-1971, fujii-1974, freund-1974, englert-1975, englert-1976, bekenstein-1980, weyl_1919, weyl_book, dicke-1962}:

\begin{align} g_{\mu\nu}\rightarrow\Omega^2g_{\mu\nu},\;\phi\rightarrow\Omega^{-1}\phi,\label{gauge-t}\end{align} where the conformal factor $\Omega=\Omega(x)$ is a positive function of the spacetime point $\Omega>0$. The transformations \eqref{gauge-t} act only on fields but neither on the spacetime points nor on the spacetime coordinates that label them. In other words; the LST \eqref{gauge-t} are not spacetime diffeomorphisms.\footnote{The conformal transformations we consider here, do not imply coordinate transformations of either kind: active or passive. Consequently, Weyl rescalings \eqref{gauge-t} do not belong in the conformal group of transformations $C(1,3)$. These can not be confounded neither with the dilatation transformations which imply coordinate transformations: $\delta x^\mu=\epsilon x^\mu$, nor with the special conformal transformations: $\delta x^\mu=2v_\nu x^\nu x^\mu-x^2v^\mu,$ which are also coordinate transformations.}

Let us consider matter fields coupled to CCSG gravity:

\begin{align} {\cal L}_\text{tot}={\cal L}_\text{ccsg}+{\cal L}_\text{mat},\label{overall-lag}\end{align} where ${\cal L}_\text{mat}={\cal L}_\text{mat}(\psi_A,\der\psi_A,g_{\mu\nu})$ is the Lagrangian density of the matter fields $\psi_A$ ($A=1, 2,..., N$). Variation of the overall Lagrangian density ${\cal L}_\text{tot}$ with respect to the metric leads to the Einstein's equation \cite{deser-1970}:

\begin{align} {\cal E}_{\mu\nu}=\frac{6}{\phi^2}T^\text{mat}_{\mu\nu},\label{einst-feq}\end{align} where the tensor ${\cal E}_{\mu\nu}$ is given by 

\begin{align} {\cal E}_{\mu\nu}\equiv&G_{\mu\nu}+\frac{6}{\phi^2}\left[\der_\mu\phi\der_\nu\phi-\frac{1}{2}g_{\mu\nu}(\der\phi)^2\right]\nonumber\\
&-\frac{1}{\phi^2}\left(\nabla_\mu\nabla_\nu-g_{\mu\nu}\nabla^2\right)\phi^2+\frac{\Lambda}{8}\phi^2g_{\mu\nu},\nonumber\end{align} and

\begin{align} T^\text{mat}_{\mu\nu}=-\frac{2}{\sqrt{-g}}\frac{\delta{\cal L}_\text{mat}}{\delta g^{\mu\nu}},\nonumber\end{align} is the stress-energy tensor (SET) of the matter fields. Above we used the following definition: $\nabla^2\equiv g^{\mu\nu}\nabla_\mu\nabla_\nu$. Variation of ${\cal L}_\text{tot}$ with respect to the scalar field yields the Klein-Gordon (KG) type equation,

\begin{align} \nabla^2\phi-\frac{\phi}{6}\,R+\frac{\Lambda}{12}\,\phi^3=0,\label{kg-feq}\end{align} where we took into account that $\delta{\cal L}_\text{mat}/\delta\phi=0$ is a vanishing identity ($0=0$), which does not modify the KG-EOM \eqref{kg-feq}. The latter is to be compared with the trace of the Einstein's EOM \eqref{einst-feq}:

\begin{align} -R-\frac{6}{\phi^2}(\der\phi)^2+\frac{3}{\phi^2}\nabla^2\phi^2+\frac{\Lambda}{2}\phi^2=\frac{6}{\phi^2}\,T^\text{mat},\nonumber\end{align} where $T^\text{mat}\equiv g^{\mu\nu}T^\text{mat}_{\mu\nu}$ is the trace of the matter SET. If take into account that $\nabla^2\phi^2=2\phi\nabla^2\phi+2(\der\phi)^2$, the above equation can be written alternatively as

\begin{align} \nabla^2\phi-\frac{\phi}{6}\,R+\frac{\Lambda}{12}\,\phi^3=\frac{1}{\phi}\,T^\text{mat}.\label{einst-feq-trace}\end{align} Direct comparison of \eqref{kg-feq} and \eqref{einst-feq-trace} leads to the well-known result that $T^\text{mat}=0$, i. e., that only massless matter fields (radiation in general) couple to gravity in this theory \cite{deser-1970}. In general it is a well-known result that if the SET of any matter field is conformal invariant in the sense that $T^\text{mat}_{\mu\nu}\rightarrow\Omega^wT^\text{mat}_{\mu\nu}$, its trace must vanish identically \cite{wald-book}.

The way proposed in \cite{deser-1970} to couple matter fields with nonvanishing masses to gravity in the CCSG is by breaking the LSS. This is achieved by introducing the self-interaction potential $V=V(\phi)$ in place of the term $\Lambda\phi^4/4!$, that is, by making the replacement: $\Lambda\phi^4/48\rightarrow V$. In this case, the trace of the Einstein-type EOM reads

\begin{align} \nabla^2\phi-\frac{\phi}{6}\,R+\frac{4}{\phi}\,V=\frac{1}{\phi}\,T^\text{mat},\nonumber\end{align} while the KG type equation is given by

\begin{align} \nabla^2\phi-\frac{\phi}{6}\,R+\der_\phi V=0,\nonumber\end{align} If we compare these equations we get that matter fields with nonvanishing masses can be consistently coupled to gravity provided that the following condition takes place:

\begin{align} \frac{4}{\phi}\,V-\der_\phi V=\frac{1}{\phi}\;T^\text{mat}.\label{diff-pot}\end{align} This requirement amounts to renounce to LSS since the only self-interaction potential which is able to preserve conformal invariance is $V\propto\phi^4,$ in which case the latter equations leads to: $T^\text{mat}=0$. It seems that LSS and nonvanishing $T^\text{mat}\neq 0$ are incompatible.

The $T^\text{mat}=0$ result also arises within the quite different context of scalar tensor theories (STT), in particular the prototype Brans-Dicke (BD) theory \cite{bd-theor}. The gravitational Lagrangian density of the BD theory reads

\begin{align} {\cal L}_\text{bd}=\frac{\sqrt{-g}}{2}\left[\Phi R-\frac{\omega}{\Phi}(\der\Phi)^2-2V\right],\label{bd-lag}\end{align} where $\Phi$, with dimensions of mass squared, stands for the BD scalar field and the constant $\omega$ is the BD coupling parameter. Under the LST \eqref{gauge-t} the BD scalar field transforms like $\Phi\rightarrow\Omega^{-2}\Phi$. The EOM derived from the overall Lagrangian ${\cal L}_\text{tot}$ in \eqref{overall-lag}, with the replacement ${\cal L}_\text{ccsg}\rightarrow{\cal L}_\text{bd}$, are the BD-Einstein's EOM \cite{quiros-rev}:

\begin{align} G_{\mu\nu}-\frac{1}{\Phi}\left(\nabla_\mu\nabla_\nu-g_{\mu\nu}\nabla^2\right)\Phi-g_{\mu\nu}\frac{V}{\Phi}&\nonumber\\
-\frac{\omega}{\Phi^2}\left[\der_\mu\Phi\der_\nu\Phi-\frac{1}{2}g_{\mu\nu}(\der\Phi)^2\right]&=\frac{1}{\Phi}\,T^\text{mat}_{\mu\nu},\label{einst-bd-feq}\end{align} and the BD-KG EOM:

\begin{align} (3+2\omega)\nabla^2\Phi=2(\Phi\der_\Phi V-2V)+T^\text{mat}.\label{kg-bd-feq}\end{align} 

The BD theory admits LSS only for the singular value of the BD coupling $\omega=-3/2$, whenever $\Phi\der_\Phi V-2V=0$ $\Rightarrow$ $V\propto\Phi^2$. Then from \eqref{einst-bd-feq} it follows that the BD theory admits local-scale symmetry only if $\omega=-3/2$ and $T^\text{mat}=0$ simultaneously, which is the same result that we have obtained in the CCSG theory of \cite{deser-1970}. However, this should not be surprising since, as can be shown by redefining the BD scalar field: $\Phi=\phi^2/6$, in this particular case ($\omega=-3/2$, $V\propto\Phi^2$,) the gravitational Lagrangian of the BD theory ${\cal L}_\text{bd}$ in \eqref{bd-lag} coincides with the CCSG Lagrangian ${\cal L}_\text{ccsg}$ in \eqref{ccs-lag}. 

In reference \cite{deser-1970} the break down of LSS is achieved through the addition of a mass term for $\phi$ in \eqref{ccs-lag}:

\begin{align} {\cal L}_\text{mass}=\frac{\sqrt{-g}}{2}\mu^2\phi^2,\label{mass-term}\end{align} where $\mu$ is a constant mass parameter. In this case $T^\text{mat}=\mu^2\phi^2$. This is consistent with the hypothesis that any mass parameter in the gravitational Lagrangian breaks down local-scale symmetry by allowing nonvanishing trace of the matter SET. This result, together with the hypothesis that LSS may not survive in the quantum domain, sets very tight constraints on the role that LSS may have played in nature.\footnote{The mildest estimate sets the lower energy bound at which conformal symmetry may have survived, at about $1$ TeV, i. e., at the scale of the break down of $SU(2)\times U(1)$ symmetry, when matter fields acquired masses. Other more pessimistic estimates establish that conformal symmetry does not survive below the grand unification or even Plack scales. At the opposite extreme, when we had energies $\sim 10^{16}-10^{19}$ GeV and higher, it is not clear the role of conformal symmetry since it is supposed that LSS does not survive in the domain where quantum gravity starts playing a role.}

The local-scale symmetry is fortunate that the above result is not definitive. For instance, if lift the mass parameter $\mu$ to a point-dependent field $\mu(x)$ which, under the conformal transformations \eqref{gauge-t}, transforms like $\phi$ does:\footnote{Recall that $\phi$ has mass dimension.} $\mu\rightarrow\Omega^{-1}\mu$, then the mass Lagrangian \eqref{mass-term} does not break LSS since, under \eqref{gauge-t}: ${\cal L}_\text{mass}\rightarrow{\cal L}_\text{mass}.$ In this case the theory ${\cal L}_\text{ccsg}+{\cal L}_\text{mass}+{\cal L}_\text{mat}$, is conformal invariant and yet, $T^\text{mat}=\mu(x)^2\phi^2\neq 0$. That the masses transform as above is not a uncommon hypothesis. For example, based on the dimensional analysis, in \cite{dicke-1962} (see \cite{faraoni-2007} for a more contemporary analysis), it is argued that the masses transform in this precise way. In reality, on the basis of dimensional analysis, it follows that the conformal transformations of the reduced Planck constant $\hbar$ and of the mass parameter $m$, are not independent of each other. The former has units $[\hbar]=[M][L]^2/[T]$, where $M$ is the mass unit. Given the transformation properties of time and length units under LST,\footnote{Under the conformal transformations \eqref{gauge-t} the units of time $[T]$ and length $[L]$ transform as the proper time element $\sqrt{-g_{00}dt^2}$, and the proper length element $\sqrt{g_{ik}dx^idx^k}$, do: $[T]\rightarrow\Omega\,[T]$ and $[L]\rightarrow\Omega\,[L]$, respectively.} we have $[L]^2/[T]\rightarrow\Omega[L]^2/[T]$, so that Planck's constant is not transformed by \eqref{gauge-t} only if the mass unit transforms like $[M]\rightarrow\Omega^{-1}[M]$, which means that the mass parameter transforms in the same way: $m\rightarrow\Omega^{-1}m$. Alternatively, if we assume that the mass parameter is unchanged by LST, then the Planck constant transforms like $\hbar\rightarrow\Omega\hbar$. 

The hypothesis that under Weyl rescaling \eqref{gauge-t} the masses transform as the scalar field $\phi$ does, can be achieved if in the Dirac Lagrangian make the replacement $m\bar\psi\psi$ $\rightarrow\kappa\phi\,\bar\psi\psi,$ where $\psi$ is the wave function (spinor) of the fermion field and $\kappa$ is a dimensionless constant \cite{hobson-2020, hobson-2022}. In an equivalent way, within the framework of the standard model (SM) of particles and fields, the mass parameter $v$ in the Higgs potential $V\propto(H^\dag H-v^2)^2$, where $H$ stands for the doublet Higgs field, must be lifted to a field \cite{bars-2014-a, nicolai-2007}: $v\rightarrow v_0\phi$, where $v_0$ is a dimensionless constant. In \cite{salam-1970} it was shown for the first time that the introduction of a real scalar or dilaton field enables one to construct conformal invariant Lagrangians for massive matter fields.

Our present work has been motivated by the hypothesis of classically unbroken conformal symmetry \cite{nicolai-2007, foot-2008, hur-2011, tang-2018, jung-2019, hu-2023}. If local scale symmetry is an unbroken symmetry in our present universe, perhaps it may explain some of the unsolved cosmological puzzles. This would require a LSI theory of gravity, for example the CCSG theory, which is the simplest generalization of general relativity. If we insist on exploring CCSG as a proper alternative for LSI gravitational theory, we have to face a difficult obstacle: How to couple arbitrary matter fields to gravity in the CCSG theory without spoiling the conformal symmetry? Otherwise, could there be another way of coupling timelike matter fields to gravity in CCSG theory than breaking conformal symmetry? 

In the present paper, we shall show that the requirement of vanishing trace of the matter SET is due to ignorance of LSS in the variational procedure and that mathematical consistency of this procedure requires incorporating conformal symmetry. Therefore, the variation of the CCSG Lagrangian \eqref{ccs-lag} with respect to the scalar field is not independent of the variation with respect to the metric. The mathematical consequence is that the KG-type EOM coincides with the trace of Einstein's type EOM, independent of whether the trace of the matter SET vanishes or not. Hence, contrary to previous claims, any matter fields (both timelike and null fields) consistently couple to gravity in the CCSG theory. The implications of this result for LSS, as well as for phenomenology, will be discussed accordingly.


The paper is organized as follows. In Section \ref{sect-mat} we show that, if appropriately consider the local scale symmetry in the variational procedure, any matter fields regardless of whether these are timelike or null fields can consistently couple to gravity in the CCSG theory. This means that vanishing trace of the matter SET is not required at all. Local-scale invariance of the classical Lagrangian density of matter fields, required for the above demonstration, is shown in Section \ref{sect-conf-mat} to be satisfied by fundamental fields such as the Proca and fermion fields, as well as by perfect fluids. In Section \ref{sect-gauge}, we expose the differences between passive and active approaches to local-scale transformations. This is a necessary discussion because only if we follow the active approach to LSTs local-scale symmetry can have phenomenological consequences. Otherwise, if we follow the passive approach, conformal invariance is a spurious (fictitious) symmetry without any physical significance. A required discussion on the fifth force arising in the CCSG theory as a consequence of coupling of timelike matter fields is given in Section \ref{sect-5-force}, while in Section \ref{sect-discu} the results of this paper are discussed and brief conclusions are given. 

Here we assume that any constants, regardless of whether they are fundamental or dimensionless (including integration constants), are not transformed by conformal transformations \eqref{gauge-t}. In addition, we adopt the following signature of the metric: $(-,+,+,+)$ and, unless otherwise stated, the units system where $\hbar=c=1,$ and $8\pi G_N=M^2_\text{pl}=1$.



\section{Coupling of arbitrary matter fields to CCSG theory}
\label{sect-mat}


From now on, we continue working with the BD scalar field $\Phi$. This means that the overall Lagrangian density ${\cal L}_\text{tot}={\cal L}_\text{ccsg}+{\cal L}_\text{mat},$ in \eqref{overall-lag} now reads

\begin{align} {\cal L}_\text{tot}=\frac{\sqrt{-g}}{2}\left[\Phi R+\frac{3}{2}\frac{(\der\Phi)^2}{\Phi}-\frac{3\Lambda}{2}\Phi^2\right]+{\cal L}_\text{mat}.\label{tot-lag}\end{align} It is not difficult to probe that, if take into account the relationship $\Phi=\phi^2/6$, the gravitational part of \eqref{tot-lag} coincides with \eqref{ccs-lag}. It is manifestly invariant under the LST:

\begin{align} g_{\mu\nu}\rightarrow\Omega^2g_{\mu\nu},\;\;\Phi\rightarrow\Omega^{-2}\Phi,\label{conf-t}\end{align} due to conformal invariance of the units: 

\begin{align} \sqrt{-g}\left[\Phi R+\frac{3}{2}\frac{(\der\Phi)^2}{\Phi}\right],\nonumber\end{align} and $\sqrt{-g}\,\Phi^2$, separately. 

In this section, in order for the overall Lagrangian \eqref{tot-lag} and the derived EOM to be invariant under the conformal transformations \eqref{conf-t}, we shall assume without proof that the Lagrangian density of the matter fields is invariant under the conformal transformations as well: ${\cal L}_\text{mat}\rightarrow{\cal L}_\text{mat}$. Then, in the next Section, we shall investigate what kind of matter fields fulfill this assumption.

The EOM which are derived by varying \eqref{tot-lag} with respect to the metric read,

\begin{align} {\cal E}_{\mu\nu}=\frac{1}{\Phi}\,T^\text{mat}_{\mu\nu},\label{mat-eom}\end{align} where, in terms of the BD scalar field, the tensor ${\cal E}_{\mu\nu}$ is given by:

\begin{align} {\cal E}_{\mu\nu}=&\,G_{\mu\nu}+\frac{3}{2\Phi^2}\left[\der_\mu\Phi\der_\nu\Phi-\frac{1}{2}g_{\mu\nu}(\der\Phi)^2\right]\nonumber\\
&-\frac{1}{\Phi}\left(\nabla_\mu\nabla_\nu-g_{\mu\nu}\nabla^2\right)\Phi+\frac{3\Lambda}{4}\Phi g_{\mu\nu}.\label{e-mn}\end{align} The trace of equation \eqref{mat-eom} reads:

\begin{align} {\cal E}=-R-\frac{3}{2}\frac{(\der\Phi)^2}{\Phi^2}+3\frac{\nabla^2\Phi}{\Phi}+3\Lambda\Phi=\frac{1}{\Phi}\,T^\text{mat},\label{mat-eom-trace}\end{align} where ${\cal E}\equiv g^{\mu\nu}{\cal E}_{\mu\nu}$ and $T^\text{mat}\equiv g^{\mu\nu}T^\text{mat}_{\mu\nu}.$ It remains to derive the EOM for the gauge field $\Phi$.



\subsection{LSS and the variational principle}


Let us show how to incorporate local-scale symmetry into the variational principle in order to obtain a consistent system of EOM, without requiring vanishing of the trace of the matter SET. 

The key point of the present discussion is to consider infinitesimal Weyl rescalings \eqref{conf-t}:

\begin{align} \delta g_{\mu\nu}=2w(x)\,g_{\mu\nu},\;\delta\Phi=-2w(x)\,\Phi,\label{conf-t-inf}\end{align} where, for convenience (and temporarily,) we have redefined the conformal factor: $\Omega(x)=e^{w(x)}$. The infinitesimal LST are of importance, for instance, to compute the Noether currents that are associated with conformal symmetry \cite{jackiw-2015}. Substituting the right-hand equation in \eqref{conf-t-inf} back into the left-hand one, we get that,

\begin{align} \delta g_{\mu\nu}=-\frac{\delta\Phi}{\Phi}g_{\mu\nu}\;\Rightarrow\;g^{\mu\nu}\delta g_{\mu\nu}+4\frac{\delta\Phi}{\Phi}=0.\label{weyl-cond}\end{align} This means that variation of the metric and of the scalar field $\Phi$ are not independent of each other, as long as local-scale symmetry takes place. 

Equivalently, as has been shown in \cite{alvarez-2015}, the Ward identity induced by conformal symmetry implies that:\footnote{See equations (2.10) and (3.5) of Ref. \cite{alvarez-2015}, for multiple and for a single scalar field, respectively}

\begin{align} g^{\mu\nu}\frac{\delta S}{\delta g^{\mu\nu}}=\Phi\frac{\delta S}{\delta\Phi},\label{noether-id}\end{align} or in terms of the variational derivative of the overall Lagrangian ${\cal L}_\text{tot}$, 

\begin{align} g^{\mu\nu}\frac{\delta{\cal L}_\text{tot}}{\delta g^{\mu\nu}}=\Phi\frac{\delta{\cal L}_\text{tot}}{\delta\Phi},\label{lag-noether-id}\end{align} where

\begin{align} \frac{\delta{\cal L}_\text{tot}}{\delta g^{\mu\nu}}=\frac{\sqrt{-g}}{2}\left[\Phi{\cal E}_{\mu\nu}-T^\text{mat}_{\mu\nu}\right],\label{tot-lag-var}\end{align} with ${\cal E}_{\mu\nu}$ given by \eqref{e-mn}. Equations \eqref{weyl-cond} and \eqref{lag-noether-id}, which have been obtained following different approaches, are different but equivalent ways of incorporating LSS into the variational procedure.

From equations \eqref{lag-noether-id} and \eqref{tot-lag-var} it follows that

\begin{align} \Phi\frac{\delta{\cal L}_\text{tot}}{\delta\Phi}=\frac{\sqrt{-g}}{2}\left(\Phi{\cal E}-T^\text{mat}\right),\label{interm-eq}\end{align} where ${\cal E}=g^{\mu\nu}{\cal E}_{\mu\nu}$ is given by \eqref{mat-eom-trace}. Hence, the variational principle $\delta{\cal L}_\text{tot}/\delta\Phi=0$, leads to the correct KG-type equation for the scalar field:

\begin{align} \nabla^2\Phi-\frac{1}{2\Phi}(\der\Phi)^2+\Lambda\Phi^2-\frac{\Phi}{3}R=\frac{T^\text{mat}}{3}.\label{mat-kg-eom}\end{align} This equation coincides with the trace of the Einstein's EOM \eqref{mat-eom}, which is given by \eqref{mat-eom-trace}, independent of whether the trace of the matter SET vanishes or not.



\subsection{Continuity equation}


Let us write the EOM \eqref{mat-eom} for arbitrary matter content in the following way: $\Phi{\cal E}_{\mu\nu}=T^\text{mat}_{\mu\nu}.$ If we take the divergence of this equation, we get,

\begin{align} \nabla^\lambda(\Phi{\cal E}_{\lambda\mu})=\frac{\der_\mu\Phi}{2}\left[-R-\frac{3}{2}\frac{(\der\Phi)^2}{\Phi^2}+3\Lambda\Phi\right.\nonumber\\
\left.+3\frac{\nabla^2\Phi}{\Phi}\right]=\nabla^\lambda T^\text{mat}_{\lambda\mu},\nonumber\end{align} where we took into account the second Bianchi identity $\nabla^\mu G_{\mu\nu}=0$, and the identity $(\nabla_\lambda\nabla_\mu-\nabla_\mu\nabla_\lambda)\nabla^\lambda\Phi=R_{\mu\lambda}\nabla^\lambda\Phi.$ If in the above equation substitute \eqref{mat-eom-trace}, we get the 'continuity' equation:

\begin{align} \nabla^\lambda T^\text{mat}_{\lambda\mu}=\frac{1}{2}\frac{\der_\mu\Phi}{\Phi}\,T^\text{mat},\label{div-set}\end{align} which is invariant with respect to the LST \eqref{conf-t}. 

Equation \eqref{div-set} is not a proper conservation equation due to the right-hand source term. This means that timelike matter fields suffer an additional non-gravitational interaction with the gauge field $\Phi,$ which should be understood as a fifth force acting only on fields with non-vanishing SET trace. For massless fields and for radiation in general with a vanishing SET trace, \eqref{div-set} is an actual conservation equation in $V_4.$


The above result entails that timelike matter fields follow worldlines that obey the following LSI equation of motion:

\begin{align} \frac{d^2x^\alpha}{ds^2}+\left\{^\alpha_{\mu\nu}\right\}\frac{dx^\mu}{ds}\frac{dx^\nu}{ds}-\frac{1}{2}\frac{\der_\mu\Phi}{\Phi}\,h^{\mu\alpha}=0,\label{tlk-worldline}\end{align} where the orthogonal projection tensor is given by

\begin{align} h^{\mu\nu}\equiv g^{\mu\nu}-\frac{dx^\mu}{ds}\frac{dx^\nu}{ds}=g^{\mu\nu}+u^\mu u^\nu,\label{h-mn}\end{align} with the fourth velocity vector $u^\mu=dx^\mu/d\tau$ ($d\tau^2=-ds^2$) satisfying the following conditions: $g_{\mu\nu}u^\mu u^\nu=-1$ and $h_{\mu\lambda}u^\lambda=0$. Equation \eqref{tlk-worldline} is not a geodesic equation in $V_4$ because of the fifth-force term $\propto\der_\mu\Phi/\Phi.$ For matter fluids the consequence of \eqref{tlk-worldline} is the continuity equation \eqref{div-set}.

In contrast, fields with vanishing mass, like photons and radiation fields in general, obey an actual conservation equation, $\nabla^\lambda T^\text{rad}_{\lambda\mu}=0,$ since their SET trace in \eqref{div-set} vanishes, $T^\text{rad}=g^{\mu\nu}T^\text{rad}_{\mu\nu}=0.$ This entails that these fields follow null geodesics of $V_4$:

\begin{align} \frac{dk^\mu}{d\xi}+\left\{^\mu_{\nu\sigma}\right\}k^\nu k^\sigma=0,\label{0-geod}\end{align} where $k^\mu\equiv dx^\mu/d\xi$ is the wave vector ($k_\mu k^\mu=0$) and $\xi$ is an affine parameter along null-geodesic.\footnote{Based on dimensional analysis it follows that $k^\mu$ has conformal weight $w=-2$, like the fourth-momentum $p^\mu=mdx^\mu/ds$ (we assume that, under the LST \eqref{conf-t}, the mass transforms like $m\rightarrow\Omega^{-1}m$ \cite{dicke-1962}). Hence, under \eqref{conf-t} the wave vector transforms like $k^\mu\rightarrow\Omega^{-2}k^\mu$ $\Rightarrow$ $d\xi\rightarrow\Omega^2d\xi$, while: $$\frac{dk^\mu}{d\xi}\rightarrow\Omega^{-4}\left[\frac{dk^\mu}{d\xi}-2k^\mu\frac{d\ln\Omega}{d\lambda}\right],$$ and $$\{^\alpha_{\lambda\nu}\}k^\lambda k^\nu\rightarrow\Omega^{-4}\left[\{^\mu_{\lambda\nu}\}k^\lambda k^\nu+2k^\mu\frac{d\ln\Omega}{d\lambda}\right],$$ so that \eqref{0-geod} is not transformed by \eqref{conf-t}.}

The following conclusion of the above analysis may come as a surprise: If we incorporate LSI into the variational principle, any matter fields -- no matter whether massless or with mass -- consistently couple to gravity in the CCSG theory \eqref{tot-lag} over $V_4$ background spaces. This is at the cost that timelike matter fields are acted on by a fifth force $\propto\der_\mu\Phi/\Phi$, while massless matter fields are transparent to this additional nongravitational interaction.\footnote{The described fifth force which acts only on timelike matter fields can not be measured in local experiments since it affects not only the motion of measured timelike fields but, also, the motion of the measuring stick, which is made of timelike particles itself. Only in experiments involving distant points in spacetime and including not only timelike matter fields but also null fields, as in redshift experiments, can the mentioned fifth force be measured.} This means that gravity in the CCSG theory with arbitrary matter content in $V_4$, is not purely geometrical since the interactions of matter are not only with the metric field but also with the non-geometric compensator field $\Phi$.



\section{Matter fields and conformal invariance}
\label{sect-conf-mat}


In the former section, to allow for conformal invariance of the overall Lagrangian \eqref{tot-lag}, we have assumed without proof that the Lagrangian density of matter fields ${\cal L}_\text{mat}$ is invariant under LST \eqref{conf-t}. Here we shall show that this assumption is satisfied by classical Lagrangians of both fundamental matter fields and background matter in the form of perfect fluids. The latter type of Lagrangian covers the most interesting cases within the context of classical gravitational interactions of matter.

An important piece of the reasoning line below is related to the point-dependent property of the mass parameter. The additional interaction of the matter fields with $\Phi$ that arises in CCSG theory may be due to the field mass being $\Phi$-dependent: $m=\kappa\sqrt\Phi,$ where $\kappa$ is some dimensionless parameter. This can be achieved if in the Dirac Lagrangian the replacement is made $m\bar\psi\psi$ $\rightarrow\kappa\sqrt{\Phi}\,\bar\psi\psi,$ where $\psi$ is the wave function (spinor) of the fermion field \cite{hobson-2020, hobson-2022}. In general, the Higgs Lagrangian of the standard model must be modified in order to respect LSS \cite{bars-2014-a, salam-1970, drechsler-1999}. Here we shall not assume any specific modification; instead we will proceed by assuming that after $SU(2)\times U(1)$ symmetry breaking, any fields $\psi_A$ acquire masses $m_A=\kappa_A\sqrt\Phi$, where the constants $\kappa_A$ are different for the different fields.


\subsection{Proca field}


Let us first consider a standard massless spin 1 field $A_\mu$, with vanishing conformal weight $w(A_\mu)=0$, so that, under the LST \eqref{conf-t} $A_\mu\rightarrow A_\mu$ ($A^\mu\rightarrow\Omega^{-2}A^\mu$) and the field strength $F_{\mu\nu}\equiv 2\nabla_{[\mu}A_{\nu]}\rightarrow F_{\mu\nu}$ is also preserved. The Lagrangian density

\begin{align} {\cal L}_\text{em}=-\frac{\sqrt{-g}}{4}\,F^2,\nonumber\end{align} where we use the notation $F^2\equiv F_{\mu\nu}F^{\mu\nu}$, is clearly conformal invariant since, under \eqref{conf-t} $\sqrt{-g}\rightarrow\Omega^4\sqrt{-g}$ and $F^2\rightarrow\Omega^{-4}F^2$. Let us add a mass term in the above Lagrangian density:

\begin{align} {\cal L}_\text{proca}=-\sqrt{-g}\left(\frac{1}{4}\,F^2+\frac{1}{2}\,m^2_P A^2\right),\label{proca-lag}\end{align} where we write $A^2\equiv A_\mu A^\mu$ for compactness. Under the LST \eqref{conf-t} $A^2\rightarrow\Omega^{-2}A^2$. Now we take into account that $m_P=\kappa_P\sqrt{\Phi}$ ($\kappa_P$ is some dimensionless constant parameter,) so that under the conformal transformations $m_P\rightarrow\Omega^{-1}m_P$. Hence, $m^2_PA^2\rightarrow\Omega^{-4}m^2_PA^2$. This demonstrates that the Proca Lagrangian density \eqref{proca-lag} is conformal invariant. Variation of the Proca Lagrangian density with respect to the metric yields the Proca stress-energy tensor:

\begin{align} T^\text{proca}_{\mu\nu}=&-\frac{2}{\sqrt{-g}}\frac{\delta{\cal L}_\text{proca}}{\delta g^{\mu\nu}}=F_{\mu\lambda}F^\lambda_{\;\;\nu}-\frac{F^2}{4}\,g_{\mu\nu}\nonumber\\
&+m^2_P\left(A_\mu A_\nu-\frac{A^2}{2}\,g_{\mu\nu}\right),\label{proca-set}\end{align} whose trace is nonvanishing: $T^\text{proca}=-m^2_PA^2$.


\subsection{Fermions}


In $V_4$ space the Dirac Lagrangian density for massless fermions reads

\begin{align} {\cal L}_\text{dirac}=\sqrt{-g}\bar\psi i\cancel{\cal D}\psi,\label{dirac-lag}\end{align} where $(\psi,\bar\psi)$ are the fermion spinor and its adjoint spinor. Both have conformal weight $w(\psi)=w(\bar\psi)=-3/2$, so that under the LST \eqref{conf-t} these transform like: $\psi\rightarrow\Omega^{-3/2}\psi$, $\bar\psi\rightarrow\Omega^{-3/2}\bar\psi$, respectively. In the above equation the derivative operator $\cancel{\cal D}:=\gamma^\mu{\cal D}_\mu$ has been introduced, where

\begin{align} {\cal D}_\mu\psi=\left[D_\mu-\frac{1}{2}\sigma_{ab}e^{b\nu}\left(\nabla_\mu e^a_\nu\right)\right]\psi,\label{cov-der-psi}\end{align} with $a,b,c=0,1,2,3$ flat spacetime indices, while $\gamma^a$ are the Dirac gamma matrices, $e^a_\mu$ are the tetrad fields such that, $g_{\mu\nu}=\eta_{ab}e^a_\mu e^b_\nu$ ($\eta_{ab}$ is the Minkowski metric). The conformal weight of the tetrads is $w(e^a_\mu)=1$ and $w(e^\mu_a)=-1$, respectively. Besides $\gamma^\mu=e^\mu_c\gamma^c$, etc. and 

\begin{align} \sigma_{ab}=\frac{1}{2}\left[\gamma_a,\gamma_b\right]=\frac{1}{4}\left(\gamma_a\gamma_b-\gamma_b\gamma_a\right),\nonumber\end{align} are the generators of the Lorentz group in the spin representation. Above we have used the standard definition of the gauge $SU(2)\times U(1)$ derivative,

\begin{align} D_\mu\psi=\left[\der_\mu+igW^i_\mu T^i-\frac{i}{2}g'YB_\mu\right]\psi,\label{su2-u1-gauge-der}\end{align} where $W^i_\mu$ and $B_\mu$ are the $SU(2)$ and $U(1)$ bosons, respectively (both have vanishing conformal weight,) while $(g,g')$ are the gauge couplings, $Y$ is the hypercharge for $\psi$ and $T^i$ are the isospin matrices.

Invariance of the Dirac Lagrangian density \eqref{dirac-lag} for massless fermions has been demonstrated in \cite{cheng-1988}. However, in the following, for completeness of the exposition, we shall give a brief and compact demonstration. First, we notice that none of the quantities with flat indices, such as, for example: $\eta_{ab}$, $\gamma^a$, $\gamma_b$, and $\sigma_{ab}$, transform under \eqref{conf-t}. In addition, the vector fields $W^i_\mu$ and $B_\mu$ are also not transformed. Second, under LST \eqref{conf-t}, we have that

\begin{align} \gamma^\mu\der_\mu\psi\rightarrow\Omega^{-5/2}\gamma^\mu\left(\der_\mu-\frac{3}{2}\der_\mu\ln\Omega\right)\psi,\label{transf-1}\end{align} while 

\begin{align} \nabla_\mu e^a_\nu\rightarrow\Omega\left(\nabla_\mu e^a_\nu-e^a_\mu\der_\nu\ln\Omega+g_{\mu\nu}e^a_\lambda\der^\lambda\ln\Omega\right),\nonumber\end{align} which leads to 

\begin{align} -&\frac{1}{2}\gamma^\mu\sigma_{ab}e^{b\nu}\nabla_\mu e^a_\nu\psi\rightarrow\nonumber\\
&\Omega^{-5/2}\gamma^\mu\left(-\frac{1}{2}\sigma_{ab}e^{b\nu}\nabla_\mu e^a_\nu+\frac{3}{2}\der_\mu\ln\Omega\right)\psi,\label{transf-2}\end{align} where in the last transformation we took into account the antisymmetry property $\sigma_{ab}=-\sigma_{ba}$ and also that $\gamma^a\sigma_{ab}=3\gamma_b/2$. Taking into account \eqref{transf-1} and \eqref{transf-2} we get that under the conformal transformations \eqref{conf-t} $\cancel{\cal D}\psi\rightarrow\Omega^{-5/2}\cancel{\cal D}\psi$, so that: $\bar\psi\cancel{\cal D}\psi\rightarrow\Omega^{-4}\bar\psi\cancel{\cal D}\psi$. This completes the demonstration that the Dirac Lagrangian density \eqref{dirac-lag} for massless fermions is conformal invariant.

Notice that adding a mass term of the fermion

\begin{align} {\cal L}_\text{dirac-mass}=\sqrt{-g}\bar\psi\left(i\cancel{\cal D}+m_\psi\right)\psi,\label{mass-dirac-lag}\end{align} does not change the above result. Actually, recalling that $m_\psi=\kappa_\psi\sqrt{\Phi}$, where $\kappa_\psi$ is some constant, so that under \eqref{conf-t}: $m_\psi\rightarrow\Omega^{-1}m_\psi$, it is easily seen that $m_\psi\bar\psi\psi\rightarrow\Omega^{-4}m_\psi\bar\psi\psi$, so that the Dirac's Lagrangian density for fermions with nonvanishing mass is also conformal invariant, ${\cal L}_\text{dirac-mass}\rightarrow{\cal L}_\text{dirac-mass}$.


\subsection{$SU(3)$ fermions and gauge bosons}


The quantum chromodynamics (QCD) $SU(3)$ symmetric Lagrangian reads,

\begin{align} {\cal L}_\text{qcd}=\sqrt{-g}\left[\bar Q\left(i\cancel{\cal D}+m_Q\right)Q-\frac{1}{4}{\cal G}^{(A)}_{\mu\nu}{\cal G}^{(A)\mu\nu}\right],\label{qcd-lag}\end{align} where the fermion's mass $m_Q=\kappa_Q\sqrt\Phi$, the gauge derivative $D_\mu$ is given by Eq. \eqref{su2-u1-gauge-der} with the addition of the term $-ig''\lambda^{(A)}{\cal G}^{(A)}_\mu/2$ ($A=1,2,...,8$), $Q$ is a triplet of spin $1/2$ fermions, $g''$ is the gauge coupling, ${\cal G}^{(A)}_\mu$ represent the gauge fields (gluons), $\lambda^{(A)}$ are the eight matrices of $SU(3)$, which obey $[\lambda^{(A)},\lambda^{(B)}]=2if_{ABC}\lambda^{(C)}$, $f_{ABC}$ are the structure constants and ${\cal G}^{(A)}_{\mu\nu}$ are the field strength tensors of the gluons,

\begin{align} {\cal G}^{(A)}_{\mu\nu}=\nabla_\mu {\cal G}^{(A)}_\nu-\nabla_\nu {\cal G}^{(A)}_\mu-g''f_{ABC}{\cal G}^{(B)}_\mu{\cal G}^{(C)}_\nu.\nonumber\end{align} 

The demonstration that the QCD Lagrangian density \eqref{qcd-lag} is a local-scale invariant object is similar to the above demonstrations, so we will not repeat it.


\subsection{Perfect fluid}


The Lagrangian density for a perfect fluid with energy density $\rho$ and barotropic pressure $p$, can be written in the following form \cite{hawking-book, faraoni_2009, berto-prd-2008}:

\begin{align} {\cal L}_\text{fluid}=-\sqrt{-g}\,\rho.\label{pfluid-lag}\end{align} Although in the bibliography one also finds that ${\cal L}_\text{fluid}=\sqrt{-g}\,p$ (see, for instance, \cite{schutz_1970}) as has been clearly shown, unless the perfect fluid couples explicitly to the curvature (which is not the case in this paper), both Lagrangian densities are equivalent \cite{faraoni_2009}. Here we assume the Lagrangian density \eqref{pfluid-lag}. 


Let us first apply the dimensional analysis to show how the energy density of a barotropic fluid transforms under the conformal transformations \eqref{conf-t}. The energy density has units: $[M]/[L]^3$, where $[M]\rightarrow\Omega^{-1}[M]$ and $[L]\rightarrow\Omega[L]$. Then $\rho\rightarrow\Omega^{-4}\rho$. (The barotropic pressure of the fluid $p$ transforms in the same way under the conformal transformations: $p\rightarrow\Omega^{-4}p$.) This means that the Lagrangian density of a perfect fluid \eqref{pfluid-lag} is conformal invariant. Applying the variational derivative to \eqref{pfluid-lag} (see \cite{hawking-book, taub_1954} for detailed application of the variational procedure to this Lagrangian) one gets that

\begin{align} \frac{\delta{\cal L}_\text{fluid}}{\delta g^{\mu\nu}}=-\frac{\sqrt{-g}}{2}\,T^\text{fluid}_{\mu\nu},\nonumber\end{align} where

\begin{align} T^\text{fluid}_{\mu\nu}=\left(\rho+p\right)u_\mu u_\nu+pg_{\mu\nu},\label{fluid-set}\end{align} stands for the fluid's SET. In order to obtain the latter equation it has to be taken into account that, an essential feature of any variational principle involving fluids, is that the world lines of the fluid's particles be among the quantities varied \cite{schutz_1970}.

Given the conformal weights of the different quantities in \eqref{fluid-set}: $w(u_\mu)=1$, $w(\rho)=w(p)=-4$ and $w(g_{\mu\nu})=2$, under the conformal transformations \eqref{conf-t} the fluid SET transforms as $T^\text{fluid}_{\mu\nu}\rightarrow\Omega^{-2}T^\text{fluid}_{\mu\nu}$, as it should be (see, for instance, \cite{dicke-1962, wald-book} or the appendix A.4 of reference \cite{wands_2000}). The same is true for any stress-energy tensor.

Let us summarize the main results of this section. We have demonstrated that under conformal transformation

\begin{align} g_{\mu\nu}&\rightarrow\bar g_{\mu\nu}=\Omega^2g_{\mu\nu},\;\Phi\rightarrow\bar\Phi=\Omega^{-2}\Phi,\nonumber\\
\Psi&\rightarrow\bar\Psi=\Omega^w\Psi,\label{conf-t'}\end{align} where $\Psi$ collectively represents the fields of matter with conformal weight $w$, the Lagrangian density of matter ${\cal L}_\text{mat}=\sqrt{-g}\,L_\text{mat}(\Psi,\der\Psi,g_{\mu\nu})$, is not transformed: ${\cal L}_\text{mat}\rightarrow{\cal L}_\text{mat}$, that is, $$\bar{\cal L}_\text{mat}(\bar\Psi,\der\bar\Psi,\bar g_{\mu\nu})={\cal L}_\text{mat}(\Psi,\der\Psi,g_{\mu\nu}).$$ Our demonstration includes classical Lagrangians of fundamental fields as well as the Lagrangian of perfect fluids, which cover the most interesting cases within the context of classical gravitational theories.

Alternatively, the main point of this section has been to show that the classical Lagrangian density of any matter field is an LSS quantity, that is, it has a conformal weight $w({\cal L}_\text{mat})=0$. Hence,

\begin{align} \frac{2}{\sqrt{-g}}\frac{\delta{\cal L}_\text{mat}}{\delta g^{\mu\nu}}\rightarrow\Omega^{-2}\frac{2}{\sqrt{-g}}\frac{\delta{\cal L}_\text{mat}}{\delta g^{\mu\nu}},\nonumber\end{align} where we took into account that $w(\sqrt{-g})=4$ and $w(g^{\mu\nu})=-2$, so that $w(\sqrt{-g}\delta g^{\mu\nu})=2$. This leads to the usual transformation property of the matter SET under the LST \cite{wald-book, dicke-1962}: $T^\text{mat}_{\mu\nu}\rightarrow\Omega^{-2}T^\text{mat}_{\mu\nu}$. 

We may conclude that the demonstration of Section \ref{sect-mat} is correct for any matter content, at least in the domain of the classical gravitational interactions of matter.


\begin{figure*}[tbh]\centering
\includegraphics[width=8cm]{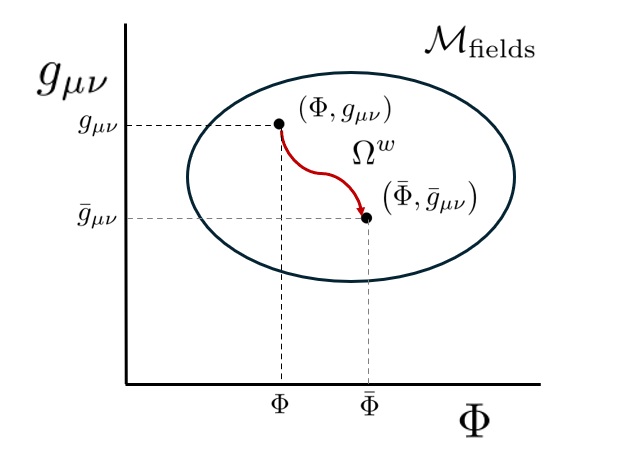}
\includegraphics[width=8cm]{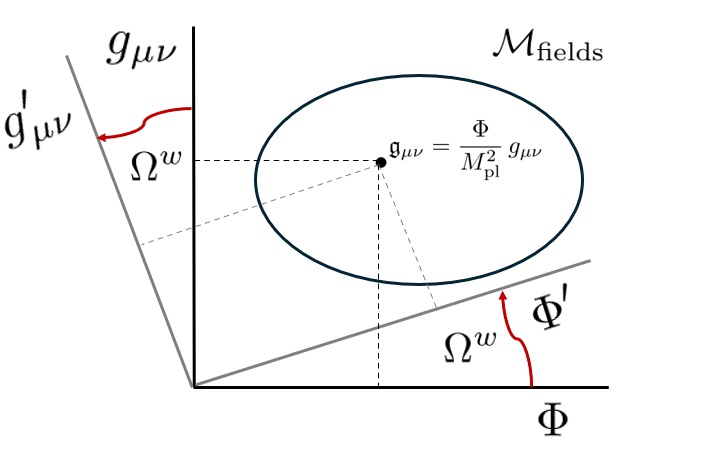}
\caption{Drawing of the field-space manifold ${\cal M}_\text{fields}$, that corresponds to the oval area in the plane $\Phi-g_{\mu\nu}.$ Each point in ${\cal M}_\text{fields}$ represents a different (vacuum) gravitational state. In the left figure the active approach to LST \eqref{conf-t} is illustrated. In this case the conformal transformations represent a ``real motion,'' i. e., a real change of the gravitational state ${\cal S}_{\bf g}:(\Phi,g_{\mu\nu})$ $\rightarrow{\bar{\cal S}}_{\bf g}:(\bar\Phi,\bar g_{\mu\nu})$. The passive standpoint on the LST is illustrated in the right figure. According to this approach the conformal transformations can be thought of as ``rotations'' of the coordinate system $R:(\Phi,g_{\mu\nu})$ in the plane $\Phi-g_{\mu\nu},$ which leaves invariant the gravitational state ${\cal S}_\mathfrak{g}:(\mathfrak{g}_{\mu\nu})$, where the invariant metric is given by \eqref{g-passive}. Here $\Omega^w$ represents the conformal factor while $w$ is the conformal weight, which is $+2$ for the metric and $-2$ for the scalar field $\Phi$.}\label{fig1}
\end{figure*}



\section{Understanding gauge fixing in CCSG theory}
\label{sect-gauge}


If we agree with statements existing in the bibliography about the physical equivalence of CCSG theory \eqref{tot-lag} and general relativity (GR), the results discussed above are meaningless, unless we clearly show the physical relevance of CCSG theory. For example, in \cite{woodard-1986} it has been stated that the gravitational Lagrangian \eqref{ccs-lag} is physically equivalent to the Einstein-Hilbert (EH) Lagrangian, for which reason there is no new physics beyond GR in the CCSG theory. This statement is correct if we follow the passive approach to conformal transformations \eqref{conf-t}. In contrast, the active approach points to the fact that CCSG theory actually bears new phenomenological consequences beyond those of GR theory.

In order to discuss passive and active conformal transformations, let us consider the geometric approach undertaken in \cite{gong_2011, karamitsos_2018}, where the scalar fields $\vphi_a$ ($a=1,2,...,n$) are treated as coordinates living in some field-space manifold, so that any transformation of the $\vphi_a$-s is then regarded as a coordinate transformation in the field space. In a similar fashion, here we assume that the metric $g_{\mu\nu}$ and the gauge scalar $\Phi$, as well as any other matter fields $\Psi$ (here $\Psi=\{\psi_1,\psi_2,...,\psi_N\}$ stands for the set of all matter fields present), are generalized coordinates in the field-space manifold: ${\cal M}_\text{fields}$. Each point in ${\cal M}_\text{fields}$ represents a gravitational state of the system.\footnote{Here, by gravitational state we understand complete knowledge of the metric $g_{\mu\nu}$ and of the compensator field $\Phi$, as well as of any matter fields $\Psi$ present, at any spacetime point. In this regard, a gravitational state can be thought of as a nonlocal concept.} 

In what follows, it will be essential to differentiate between different gravitational states in ${\cal M}_\text{fields}$; $S:(g_{\mu\nu},\Phi,\Psi),$ $\bar S:(\bar g_{\mu\nu},\bar\Phi,\bar\Psi),$ $\bar{\bar S}:(\bar{\bar g}_{\mu\nu},\bar{\bar\Phi},\bar{\bar\Psi}),$ etc. and different representations; $R:(g_{\mu\nu},\Phi,\Psi)$, $R':(g'_{\mu\nu},\Phi',\Psi'),$ $R'':(g''_{\mu\nu},\Phi'',\Psi''),$ etc. of the same gravitational state $S$ in ${\cal M}_\text{fields}$. This will allow us to differentiate passive and active LST (PLST and ALST, respectively) in the space of fields. This is illustrated in FIG. \ref{fig1} for the vacuum case. In the figure, the field-space manifold is represented by an oval area in the $\Phi-g_{\mu\nu}$ plane. The points in ${\cal M}_\text{fields}$, which have coordinates $\Phi$ and $g_{\mu\nu}$, represent different vacuum gravitational states and therefore have different phenomenological consequences. The active approach is illustrated in the left panel of FIG. \ref{fig1}. ALSTs can be understood as an actual motion in ${\cal M}_\text{fields}$, i. e. as a real change of the vacuum gravitational state. Meanwhile, as illustrated in the right panel of the figure, PLST may be thought of as ``rotations'' of the coordinate system $X^I=(\Phi,g_{\mu\nu})$ in field space.


Since passive transformations are merely different representations of the same physical state, physically meaningful quantities must be invariant under \eqref{conf-t}. This implies, in particular, that the physically meaningful metric is the conformal invariant tensor:

\begin{align} \mathfrak{g}_{\mu\nu}\equiv\frac{\Phi}{M^2_\text{pl}}\,g_{\mu\nu};\;\mathfrak{g}'_{\mu\nu}=\mathfrak{g}_{\mu\nu}\;\Rightarrow\;\Phi'g'_{\mu\nu}=\Phi g_{\mu\nu},\label{g-passive}\end{align} so that $g_{\mu\nu}$ and $\Phi$, are auxiliary fields which suffer the PLST \eqref{conf-t}, while the physically meaningful line element reads: $d\mathfrak{s}^2=\mathfrak{g}_{\mu\nu}dx^\mu dx^\nu$. In a similar way the physically meaningful matter fields can be represented by the invariant quantities: $\pi_A\equiv\left(\Phi/M_\text{pl}\right)^\frac{\omega_A}{2}\psi_A\;\Rightarrow\;\pi'_A=\pi_A$.


If we follow the passive approach to LST the Weyl rescalings \eqref{conf-t} collapse to the identity transformations:\footnote{The physical curvature invariants, such as the curvature scalar and the Kretschmann invariant of the physical metric, are given by the following expressions. $$\mathfrak{R}=\mathfrak{g}^{\mu\nu}\mathfrak{R}_{\mu\nu},\;K(\mathfrak{g}):=\mathfrak{R}^{\sigma\mu\lambda\nu}\mathfrak{R}_{\sigma\mu\lambda\nu},$$ where $\mathfrak{R}_{\mu\nu}=\mathfrak{g}^{\lambda\sigma}\mathfrak{R}_{\lambda\mu\sigma\nu}$ is the Ricci tensor and $\mathfrak{R}^\alpha_{\;\;\mu\beta\nu}$ is the Riemann-Christoffel curvature tensor of the physical metric $\mathfrak{g}_{\mu\nu}.$ These quantities are defined with respect to the affine connection: $$\mathfrak{C}^\alpha_{\;\;\mu\nu}:=\frac{1}{2}\mathfrak{g}^{\alpha\lambda}\left(\der_\nu\mathfrak{g}_{\mu\lambda}+\der_\mu\mathfrak{g}_{\nu\lambda}-\der_\lambda\mathfrak{g}_{\mu\nu}\right),$$ which coincides with the Levi-Civita (LC) connection of the physical metric.} $\mathfrak{g}_{\mu\nu}\rightarrow\mathfrak{g}_{\mu\nu}$ ($\sqrt{-\mathfrak{g}}\rightarrow\sqrt{-\mathfrak{g}}$), $\mathfrak{C}^\alpha_{\;\;\mu\nu}\rightarrow\mathfrak{C}^\alpha_{\;\;\mu\nu}$, $\mathfrak{R}^\alpha_{\;\;\mu\beta\nu}\rightarrow\mathfrak{R}^\alpha_{\;\;\mu\beta\nu}$, $\mathfrak{R}_{\mu\nu}\rightarrow\mathfrak{R}_{\mu\nu}$ and $\mathfrak{R}\rightarrow\mathfrak{R}$, which means full trivialization of the LST.\footnote{Let us assume from start that $\mathfrak{g}_{\mu\nu}$ is the physical metric tensor in a gravitational theory with given symmetries, including the spacetime diffeomorphisms. Let us further introduce the auxiliary fields without independent physical meaning: $g_{\mu\nu}$ and $\Phi$, in such a way that $\mathfrak{g}_{\mu\nu}\equiv(\Phi/M^2_\text{pl})g_{\mu\nu}$. Any transformation of the auxiliary fields that leaves the physical metric invariant, as in the PSLT, is a spurious or fictitious symmetry of the system. This result aligns with the demonstration in \cite{jackiw-2015} (see also \cite{oda-2022, rodrigo-arxiv}) that the Noether current corresponding to LSS is identically vanishing, so that the PLST are ``fake transformations''.} This is what the authors of \cite{woodard-1986} have demonstrated. Actually, if we rewrite the gravitational piece of the Lagrangian \eqref{tot-lag} in terms of the physical metric $\mathfrak{g}_{\mu\nu}$ and its derivatives, we get the EH Lagrangian density:

\begin{align} {\cal L}_\text{eh}=\frac{\sqrt{-\mathfrak{g}}\,M^2_\text{pl}}{2}\left\{\mathfrak{R}-2\Lambda_0\right\},\label{eh-lag}\end{align} where $\Lambda_0=3\Lambda M^2_\text{pl}/4$ is the cosmological constant.


The above analysis shows that the PLSTs are not an actual symmetry of the theory because these act on auxiliary fields without independent physical meaning. In contrast, ALST relate different gravitational states in field space ${\cal M}_\text{fields}$: $S:(g_{\mu\nu},\Phi,\Psi)$, $\bar S:(\bar g_{\mu\nu},\bar\Phi,\bar\Psi)$, $S\neq\bar S$. This leads, for instance, to different curvature invariants, etc. In this case, the metric $g_{\mu\nu}$ and the gauge scalar $\Phi$, have an independent physical meaning. 





In this paper, we adopt the ALST approach, as it is the only way in which both the known results in the bibliography \cite{deser-1970, englert-1975, kallosh_1975, fradkin-1978, antoniadis-1984}, as well as the results discussed in previous sections, make sense. Adoption of the active approach to conformal transformations \eqref{conf-t} means that local-scale symmetry has phenomenological consequences.


\subsection{Gauge fixing and phenomenological impact of LSS}


The CCSG theory is mathematically expressed by the Lagrangian density \eqref{tot-lag} plus the derived EOM \eqref{mat-eom}, \eqref{mat-kg-eom} and the continuity equation \eqref{div-set}. The fact that the KG-type EOM \eqref{mat-kg-eom} coincides with the trace \eqref{mat-eom-trace} of the Einstein-type EOM \eqref{mat-eom}, implies that either one of the metric functions $g_{\mu\nu}(x)$ or the scalar field $\Phi(x)$, is a free function. In this paper, for definiteness, we assume that $\Phi(x)$ is a free function since it does not obey an independent EOM. This is an inevitable consequence of local scale invariance, since, in addition to the four degrees of freedom to make diffeomorphisms, an additional degree of freedom is required to make conformal transformations \cite{anderson-1971}. The freedom to choose any function $\Phi=\Phi(x)$ we want is what we call the gauge freedom, and the scalar field $\Phi$ is then called the gauge field.

The next required step is to realize that since we follow the ALST approach (see above), every gauge has an associated phenomenology which is different from the phenomenology carried by any other gauge. In what follows, without loss of generality, we will identify a gauge with a gravitational state in the field space ${\cal M}_\text{fields}$; gauge $\Leftrightarrow$ $S:(g_{\mu\nu}(x),\Phi(x),\Psi(x))$. For illustration, fix two different gauges, say ``gauge $i$'' and ``gauge $j$'': $S_i:(g^{(i)}_{\mu\nu},\Phi_i,\Psi_i)$ and $S_j:(g^{(j)}_{\mu\nu},\Phi_j,\Psi_j)$, respectively. Both gauges obey the same EOM \eqref{mat-eom}, \eqref{div-set} and are related by the following conformal transformations:

\begin{align} &g^j_{\mu\nu}=\Omega^2_{ji}g^i_{\mu\nu},\;\Phi_j=\Omega^{-2}_{ij}\Phi_i,\nonumber\\
&\Psi_j=\Omega^{w_A}_{ji}\Psi_i\;\Rightarrow\;\psi^j_A=\Omega^{w_A}_{ji}\psi^i_A,\nonumber\end{align} or if we realize that $\Omega_{ij}=\sqrt{\Phi_i/\Phi_j}$ then

\begin{align} g^j_{\mu\nu}=\frac{\Phi_i}{\Phi_j}\,g^i_{\mu\nu},\;\Psi_j=\left(\frac{\Phi_i}{\Phi_j}\right)^\frac{w_A}{2}\Psi_i.\label{ij-conf-t}\end{align} In the above equations $i, j$ are just labels (either numbers, letters or even names) so that these should not be confounded with tensor indices. This means, in particular, that repeated labels do not entail summation.


\subsubsection{GR gauge}


A particularly interesting gauge is when $\Phi=\Phi_0=$ const. If we set $\Phi_0=M^2_\text{pl}$, we call this the ``GR gauge''. We write $S_\text{gr}:(g^\text{gr}_{\mu\nu},\Psi_\text{gr})$. The EOM resulting from the substitution of $\Phi=M^2_\text{pl}$ in \eqref{mat-eom} and \eqref{div-set} are ($\Lambda_0\equiv 3\Lambda M^2_\text{pl}/4$)

\begin{align} G_{\mu\nu}+\Lambda_0 g_{\mu\nu}=\frac{1}{M^2_\text{pl}}\,T^\text{mat}_{\mu\nu},\;\nabla^\lambda T^\text{mat}_{\lambda\mu}=0,\nonumber\end{align} which are the GR equations of motion and the conservation equation, respectively.

According to \eqref{ij-conf-t} the GR gauge $S_\text{gr}:(g^\text{gr}_{\mu\nu},\Psi_\text{gr})$, is related to any other (arbitrary) gauge $S:(g_{\mu\nu},\Phi,\Psi)$ of the CCSG theory by conformal transformation,

\begin{align} g_{\mu\nu}=\frac{M^2_\text{pl}}{\Phi}\,g^\text{gr}_{\mu\nu},\;\Psi=\left(\frac{M^2_\text{pl}}{\Phi}\right)^\frac{w_A}{2}\Psi_\text{gr}.\nonumber\end{align} This means that: 1) GR with its associated phenomenology is one of the infinitely many possible outcomes of the CCSG theory and 2) any solution that is conformal to any GR solution has the same chance of describing gravitational phenomena as the GR solution itself.

Contrary to known arguments in the bibliography that state that there is no new phenomenology behind that of GR in CCSG theory \cite{woodard-1986}, according to the active approach to conformal transformations which is the one assumed in this paper, any gauge of CCSG theory and its associated phenomenology can be a potential alternative to explain given gravitational phenomena. Only the experiment is able to pick out a gauge among the infinity of them, which is the one that better explains the experimental evidence.


\section{The fifth force}
\label{sect-5-force}


A ``unwanted'' consequence of consistent coupling of timelike matter fields to gravity in the CCSG theory over $V_4$ space is the arising of a fifth force term $f^\mu$ in the non-geodesic EOM of timelike particles \eqref{tlk-worldline} that is given by

\begin{align} f^\mu=\frac{1}{2}\,h^{\mu\lambda}\frac{\der_\lambda\Phi}{\Phi}.\label{5-force}\end{align} It leads to the continuity equation \eqref{div-set} being nonhomogeneous conservation equation:

\begin{align} \nabla^\lambda T^\text{mat}_{\lambda\mu}=\frac{1}{2}\frac{\der_\mu\Phi}{\Phi}\,T^\text{mat}.\nonumber\end{align} 

From \eqref{5-force} it follows that $u_\lambda f^\lambda=0$, that is, the fifth force is orthogonal to the four-velocity of the timelike particle. Meanwhile, from the above continuity equation it follows that the additional non-gravitational interaction does not affect the radiation and null fields. Both properties of the fifth force above \eqref{5-force}, are shared by other similar additional interactions found in the bibliography. In \cite{berto-prd-2007}, for example, $f(R)$ theories of gravity with a direct coupling between matter and a function of the Ricci scalar are investigated. The arising of an extra force with the same properties as \eqref{5-force} is confirmed. The corresponding acceleration law is obtained in the weak-field limit, and connections with MOND and the Pioneer anomaly are further discussed, leading to the extra force being an alternative explanation of the dark matter. However, in \cite{faraoni-cqg-2008}, based on the analysis of the motion of matter fields and its implications for the equivalence principle, this result was concluded to be questionable.\footnote{Even if the fifth force \eqref{5-force} cannot explain the dark matter problem in the way proposed in \cite{berto-prd-2007}, a static spherically symmetric solution of the vacuum EOM \eqref{mat-eom}: ${\cal E}_{\mu\nu}=0$, was found \cite{harko-epjc-2022}, which allowed one to explain the galaxies rotation curves \cite{harko-pdu-2024}.}


One of the main objections against the fifth force is due to the existing very strict observational constraints \cite{will-lrr-2014}. In the present case, these constraints set limits on the behavior of the gauge field $\Phi$ in the solar system. For example, one may infer that in the neighborhood of the solar system $\Phi\approx M^2_\text{pl}$ is almost a constant, while in cosmological scales it may be a function of cosmic time $t$: $\Phi=\Phi(t)$. Hence, the extra force may be related with accelerated expansion without the need for dark energy. Let us take a Friedmann-Robertson-Walker (FRW) metric: $ds^2=-dt^2+\delta_{ij}dx^i dx^j$, where $i,j=1,2,3$ are spatial indices. The independent EOM \eqref{mat-eom} and \eqref{div-set}, are the Friedmann equation

\begin{align} \left(H+\frac{1}{2}\frac{\dot\Phi}{\Phi}\right)^2=\frac{\rho}{3\Phi},\label{fried-eq}\end{align} and the continuity equation

\begin{align} \dot\rho+3H\rho=\frac{1}{2}\frac{\dot\Phi}{\Phi}\,\rho,\label{cont-eq}\end{align} respectively, where for simplicity we have omitted the $\Lambda$-term, the dot means derivative with respect to the cosmic time $t$ and, as usual, $H\equiv\dot a/a$. We consider dust fluid, so that the barotropic pressure vanishes $p=0$. Notice that the Raychaudhuri equation coming from ${\cal E}_{ij}=T^\text{mat}_{ij}/\Phi$, is not an independent equation since it can be obtained by taking the $t$-derivative of \eqref{fried-eq} and considering \eqref{cont-eq}.


The integration of the continuity equation \eqref{cont-eq} leads to $\rho=k\sqrt\Phi/a^3$, where $k$ is an integration constant. The resulting conformal-invariant EOM reads:

\begin{align} \left(\frac{v'}{v}\right)^2=\frac{k}{3v},\label{lsi-eom}\end{align} where we have introduced the new variable $v=a\sqrt\Phi$ and the prime denotes derivative with respect to the conformal time $\tau=\int dt/a$. Integration of \eqref{lsi-eom} leads to

\begin{align} v=a(\tau)\sqrt{\Phi(\tau)}=\frac{k}{12}(\tau-\tau_0)^2,\label{sol-fried-eq}\end{align} where $\tau_0$ is another integration constant. As seen most we can determine from the cosmological EOM of the CCSG theory is the LSI combination $a\sqrt\Phi$. Either $a=a(\tau)$ or $\Phi=\Phi(\tau)$ must be arbitrarily fixed (gauge fixing). For instance, let us fix $a$ to reproduce the observed cosmic dynamics, then $\Phi$ is determined through \eqref{sol-fried-eq}: $\Phi(\tau)=k(\tau-\tau_0)^2/12a(\tau)$. Then, since according to the active approach to LSS any given gauge choice leads to its own phenomenology which is different from the phenomenology of any other gauge, only the gauge whose $a(\tau)$-dynamics fit the cosmological observations, is the one that better describes our universe. This is the way in which the problem with the present accelerated stage of cosmic expansion is solved in the CCSG theory.

Whether or not dark matter and dark energy issues can be explained as manifestations of local-scale symmetry within the framework of CCSG theory, is a question that deserves further investigation.



\section{Discussion and conclusion}
\label{sect-discu}


The association of traceless matter SET with LSS within the context of conformal invariant gravitational theories has been undertaken as a fait accompli over the past several decades. This poses significant limitations to LSI gravitational theories since only matter fields with vanishing mass (radiation, in general) can couple to gravity. A solution to this issue was proposed in \cite{deser-1970} considering an LSS breaking self-interaction potential \eqref{diff-pot} for the scalar field. If we wish to preserve LSS the latter solution is not adequate. An alternative solution was then proposed in \cite{anderson-1971} where it is proposed to replace $g_{\mu\nu}$ in the matter action everywhere by the conformal invariant product $\mathfrak{g}_{\mu\nu}=\Phi g_{\mu\nu}$. In this case $\mathfrak{g}_{\mu\nu}$ must be considered as the physical metric (the one to which the matter fields are coupled). But then the physical curvature quantities are to be given in terms of the physical metric and of its derivatives. This leads the resulting theory \eqref{eh-lag} to be just Einstein's GR theory. In addition, as discussed in section \ref{sect-gauge}, the conformal transformations \eqref{conf-t} act on the now auxiliary fields $g_{\mu\nu}$ and $\Phi$, but not on the physical metric $\mathfrak{g}_{\mu\nu}$ that is conformal invariant. This means that LSS is not an actual symmetry, but a spurious one. 


In this paper, we have developed an approach to local-scale symmetry where the latter can be an actual symmetry at least in the domain of classical gravitational interactions of matter. Our approach is based on the following requirements. 


{\bf 1.- Gravitational coupling of any matter fields.} In section \ref{sect-mat} we have shown how to couple matter fields to gravitation in the CCSG theory \eqref{tot-lag} over $V_4$ space without requiring vanishing of the trace of matter SET and without renouncing to local-scale symmetry. This has been possible due to the proper consideration of LSS in the variational procedure. 

This finding shows that the very well-known constraint $T^\text{mat}=g^{\mu\nu}T^\text{mat}_{\mu\nu}=0,$ which is often associated with conformal symmetry and conformal invariant gravity theories, is just an artifact of incorrectly assuming that the variations of the overall Lagrangian with respect to the metric and with respect to the gauge field $\Phi$, are independent of each other. 


{\bf 2.- LSI matter EOM.} An important assumption in the above demonstration is that the Lagrangian density of the matter fields ${\cal L}_\text{mat}$ is itself conformal invariant. Since this is not a trivial requirement, in Section \ref{sect-conf-mat}, we investigate whether this is a correct assumption. We have demonstrated that, at least for classical Lagrangians of fundamental fields, as well as for the Lagrangian of a perfect fluid, the assumption is correct. In other words, we demonstrated that under the conformal transformations \eqref{conf-t'} ${\cal L}_\text{mat}(\Psi,\der\Psi,g_{\mu\nu})\rightarrow\bar{\cal L}_\text{mat}(\bar\Psi,\bar\der\Psi,\bar g_{\mu\nu})={\cal L}_\text{mat}(\Psi,\der\Psi,g_{\mu\nu})$. As an illustration, take Proca Lagrangian \eqref{proca-lag}. It can be easily seen that, since $\sqrt{-\bar g}=\Omega^4\sqrt{-g}$, $\bar F^2=\Omega^{-4}F^2$, $\bar A^2=\Omega^{-2}A^2$ and $\bar m_P=\Omega^{-1}m_P$,

\begin{align} \bar{\cal L}_\text{proca}&=-\sqrt{-\bar g}\left(\frac{1}{4}\,\bar F^2+\frac{1}{2}\,\bar m^2_P\bar A^2\right)\nonumber\\
&=-\sqrt{-g}\left(\frac{1}{4}\,F^2+\frac{1}{2}\,m^2_PA^2\right)={\cal L}_\text{proca}.\nonumber\end{align}


{\bf 3.- Active approach to LST.} The remaining important piece of our approach to CCSG theory is the assumption of the ALST standpoint. The result is that each gauge in the infinite class of conformal invariance of the CCSG theory carries a proper phenomenology. This means that choosing a gauge has phenomenological consequences, unlike the standard PLST viewpoint, according to which choosing a specific gauge is physically meaningless.


The resulting conformal-invariant formalism can explain the classical gravitational interactions of both time-like and null matter fields. This is at the cost that timelike fields are affected by a non-gravitational fifth force. 


The major achievement of the present investigation has been to confirm the hypothesis previously proposed in \cite{nicolai-2007, foot-2008, hur-2011, tang-2018, jung-2019, hu-2023} that LSS can be a classical (unbroken) symmetry of gravitational laws. This opens up the way for this symmetry to be a possible alternative explanation of (at least some of) present cosmological issues such as dark matter and dark energy problems.


{\bf Acknowledgments} The author acknowledges FORDECYT-PRONACES-CONACYT for supporting the present research under grant CF-MG-2558591.  





\end{document}